\documentstyle[12pt,epsf]{article}

\if@twoside
 \else
 \fi

 \parskip \medskipamount
\begin{document}
\title{Sphaleron transitions in a realistic heat bath}

\author{A. Krasnitz\thanks{Supported by the Swiss National Science Foundation.}
\\
        IPS, ETH-Zentrum, CH8092 Zurich, Switzerland\\
        and\\
        R. Potting\\
        Universidade do Algarve\\
        Unidade de Ci\^encias Exactas e Humanas,\\
        Campus de Gambelas, 8000 Faro, Portugal}
\date{August 1993}
\maketitle

\begin{abstract}
We measure the diffusion rate of Chern-Simons number in the
(1+1)-dimensional Abelian Higgs model interacting with a realistic heat bath
for temperatures between 1/13 and 1/3 times the sphaleron energy. It is found
that the measured rate is close to that predicted by one-loop calculation
at the lower end of the temperature range considered but falls at least an
order
of magnitude short of one-loop estimate at the upper end of that range. We show
numerically that the sphaleron approximation breaks down as soon as the
gauge-invariant two-point function yields correlation length close to the
sphaleron size.
\end{abstract}
\vspace{-17cm}
\begin{flushright}
IPS Research Report No. 93-10
\end{flushright}
\newpage
Anomalous electroweak baryon-number violation may have played an
important role in setting the baryon number of the Universe to its present
value \cite{Shaplat91,ML}. At temperatures above the gauge-boson mass scale
electroweak
baryon-number nonconservation is dominated by hopping over the finite-energy
barriers separating topologically distinct vacua of the bosonic sector.
Determination of the corresponding transition rate is a challenging
nonperturbative problem, even in the range of validity of the classical
approximation. At the lower temperature end of that range the barrier crossings
are likely to occur in the vicinity of the saddle point (known as a sphaleron)
of the energy functional. The rate can then be estimated using a
field-theoretic
extension of transition-state theory (TST) \cite{Langer,HTB}. At higher
temperatures this analytic tool is no longer
available, and direct measurement of the rate in real-time numerical
simulations
of a lattice gauge-Higgs system is the only remaining possibility.

Analytical saddle-point estimates of the rate in the Standard Model are
complicated by the fact that the corresponding sphaleron field configuration
is not known exactly. At the
same time, numerical real-time simulations of that system in its
low-temperature regime carry an enormous computational cost and are yet to be
performed \cite{Shaplat91,Amb3d}. In this situation lower-dimensional models
become a very useful
test ground on which activation-theory predictions can be confronted by
numerical experiments \cite{BdFH,us,BTs,SB,BdFL,GRS,GS,GST,Tang,MW,KM,Dine}.
For this reason the (1+1)-dimensional
Abelian Higgs model (AHM) studied numerically in this work has attracted much
attention recently \cite{GRS,BdFH,BTs,GST,Tang}.

Determination of the transition rate should include its proper averaging over
the canonical ensemble in the phase space of a system in question. One way to
achieve that would be to generate the canonical ensemble of initial
configurations, subject each of these configurations to the Hamitonian
evolution, and average the transition rate over the initial states of the
system. Such procedure, while being perfectly valid, is very costly
computationally. To date, a single or a small number of initial
configurations have been used in Hamiltonian simulations of AHM
\cite{GRS,Tang}.
In addition, preparing initial configurations in case of a gauge theory
presents a technical difficulty: if a standard importance-sampling method is
used, resulting configurations will in general violate Gauss' law. A special
cooling procedure is required to eliminate static charge \cite{GRS}. It is not
clear that such cooling does not cause the sample to deviate from the intended
canonical ensemble.

Another way to obtain the canonical ensemble average of the rate is to
replace Hamiltonian evolution by evolution in a heat bath. The simplest form of
the latter is implemented using phenomenological Langevin equations of motion.
While Langevin approach guarantees thermalization of the system, it does so at
the expense of introducing an artificial viscosity parameter, thereby
altering bulk dynamical properties of a field-theoretic system.
Numerical studies performed on different models show that transition rates
indeed strongly depend on viscosity \cite{SB,BdFL}. Recent analytical work
\cite{Boyan2} has also shown the impact of heat-bath properties on quantities
of transition-rate type. In case of AHM, there also is a technical difficulty
with the conventional Langevin approach: in order to maintain gauge invariance,
one is forced to use polar coordinates for the Higgs field; whenever the latter
vanishes, the equations of motion are singular \cite{BdFH,Tang}.

Recently we have proposed and tested a new method in which a
field-theoretic system interacts with a heat bath at its boundaries \cite{us}.
The heat bath is constructed so as to imitate an infinite extension of the
system beyond the boundaries. Technically this means that the fields in the
bulk
of the system evolve according to the Hamiltonian equations of motion, while
boundary fields are subject to Langevin evolution with a non-Markovian friction
kernel and colored noise. Our construction approximates a natural situation
in which open systems are immersed in a {\it similar} environment. In this way,
dynamical evolution and canonical ensemble averaging occur at the same time
while bulk dynamical properties of the system are intact. Moreover, the new
procedure does not suffer from the technical difficulties of the two old ones.
In this work we apply the realistic heat bath (RHB) method to the study of
sphaleron transitions in AHM.

Earlier real-time simulations of AHM \cite{GRS,BdFH} found that at low
temperatures (about 0.1 of the sphaleron energy) the temperature dependence of
the rate qualitatively agrees with that predicted by the 1-loop (TST)
calculation of Ref. \cite{BTs}. The temperature range of our simulation is
wider and includes somewhat higher temperatures, up to about 1/3 the sphaleron
energy, for which the TST result may no longer be reliable. This allows us to
estimate the temperature at which TST loses validity. As will be demonstrated
in the following, this is the temperature at which the scalar field correlation
length falls below the linear size of the sphaleron. It is also interesting
to determine the sign and magnitude of the rate deviation from the TST
prediction. If the rate we measure falls considerably short of the latter,
it might be indicative of the entropic rate suppression which reflects the
difficulty of creating a coherent configuration in high-temperature plasma.
Our results, presented in the following, do indeed show
dramatic slowdown of the rate growth.

Our starting point is the (1+1)-dimensional lattice AHM Lagrangian which in
suitably chosen units reads \cite{BdFH}
\begin{eqnarray}
L&=&{a\over2}\sum_j\biggl[
{1\over\xi}(\dot A^1_{j,j+1}-{A^0_{j+1}-A^0_{j}\over a})^2+
|(\partial_0-iA^0_j)\phi_j|^2\cr
&&\qquad{}-a^{-2}|\phi_{j+1}-\exp\left(iaA^1_{j,j+1}\right)\phi_i|^2
-{1\over 2}\left(|\phi_j|^2-1\right)^2\biggr].
\end{eqnarray}
Here $j$ labels sites of a chain whose lattice spacing is $a$. The temporal
component of a vector potential, $A^0_j$, and a complex scalar field $\phi_j$
reside on sites of the chain, whereas the spatial component of the vector
potential, $A^1_{j,j+1}$, resides on links. Imposing the $A^0=0$ condition
one obtains a Hamiltonian (we shall drop the Lorentz index of $A^1$ from now
on)
\begin{equation}
H={a\over 2}\sum_j\left[\left({{\xi E_{j,j+1}}\over a}\right)^2
+|{\pi_j\over a}|^2
+|\phi_{j+1}-\exp\left(iaA_{j,j+1}\right)\phi_j|^2
+{1\over 2}\left(|\phi_j|^2-1\right)^2\right],
\label{hxy}
\end{equation}
where $\pi_j$ and $E_{j,j+1}$ are canonically conjugate momenta of $\phi_j$
and $A_{j,j+1}$, respectively. The Hamiltonian equations of motion
\begin{eqnarray}
\dot A_{j,j+1}&=&{\xi\over a}E_{j,j+1},\nonumber\\
\dot E_{j,j+1}&=&i\phi_j\left(\exp(iaA_{j,j+1})\phi^*_{j+1}-\phi^*_j\right)+
{\rm h.c.},\nonumber\\
\dot\phi_j&=&{1\over a}\pi^*_j,\nonumber\\
\dot\pi_j&=&{1\over a}\left(\exp(iaA_{j,j+1})\phi^*_{j+1}+
\exp(-iaA_{j-1,j})\phi^*_{j-1}-2\phi^*_j\right)\nonumber\\
&&-a\phi^*_j\left(|\phi_j|^2-1\right)
\label{eqmoxy}
\end{eqnarray}
are supplemented by the Gauss' law constraint
\begin{equation}
{1\over a}\left(E_{j,j+1}-E_{j-1,j}\right)={\rm Im}\left(\pi_j\phi^*_j\right).
\label{glaw}
\end{equation}
The same dynamics can be described in terms of gauge-invariant variables.
To this end, the Higgs field is
rewritten in polar coordinates: $\phi_j=\rho_j\exp(i\alpha_j)$. Defining
$b_{j,j+1}=\alpha_{j+1}-\alpha_j-aA_{j,j+1}$,
$\epsilon_{j,j+1}={1\over a}E_{j,j+1}$ and introducing canonical momentum
$\pi^\rho_j$ for $\rho_j$ we see that (\ref{eqmoxy}) together with (\ref{glaw})
is equivalent to
\begin{eqnarray}
\dot\epsilon_{j,j+1}&=&{1\over a}\rho_j\rho_{j+1}\sin b_{j,j+1},\nonumber\\
\dot b_{j,j+1}&=&{1\over a}\left({{\epsilon_{j+1,j+2}-\epsilon_{j,j+1}}
\over\rho_{j+1}^2}-{{\epsilon_{j,j+1}-\epsilon_{j-1,j}}\over\rho_j^2}\right)
-a\xi\epsilon_{j,j+1},\nonumber\\
\dot\rho_j&=&{1\over a}\pi^\rho_j,\nonumber\\
\dot\pi^\rho_j&=&{{(\epsilon_{j,j+1}-\epsilon_{j-1,j})}\over{a\rho_j^3}}
+{1\over a}\left(\rho_{j+1}\cos b_{j,j+1}+\rho_{j-1}\cos b_{j-1,j}
-2\rho_j \right)\nonumber\\
&&-a\rho_j\left(\rho_j^2-1\right).
\label{eqmopol}
\end{eqnarray}
The equations of motion in this
form involve only two pairs of real canonical variables, namely,
$\rho_j,\pi^\rho_j$ and $\epsilon_{j,j+1},b_{j,j+1}$. It is easy to see that
(\ref{eqmopol}) follow from the Hamiltonian
\begin{eqnarray}
H'&=&{a\over 2}\sum_j
\left(\xi\epsilon_{j,j+1}^2
+\left({{\epsilon_{j,j+1}-\epsilon_{j-1,j}}\over{a\rho_j}}\right)^2
+\left({\pi^\rho_j\over a}\right)^2+{2\over a^2}\left(\rho_j^2-
\rho_j\rho_{j+1}\cos b_{j,j+1}\right)\right)\nonumber\\
&&+{a\over 4}\sum_j\left(\rho_j^2-1\right)^2
\label{hpol}
\end{eqnarray}
obtained from polar-coordinate form of (\ref{hxy}) by substituting
(\ref{glaw}).

In the following we shall use both presented forms of the equations of motion.
On one hand, the Cartesian form (\ref{eqmoxy}) allows better numerical handling
of sphaleronlike
field configurations in which the Higgs field is close to zero at one or
more sites. For this reason we use it for real-time evolution in the bulk of
an open gauge-Higgs system. On the other hand, the gauge-invariant form
(\ref{eqmopol}) involves less degrees of freedom and lends itself easier to
linearization. We therefore use it for the heat bath construction.

As a heat bath we take AHM linearized in the vicinity of one of its
gauge-equivalent vacua, which, as is well known,
is a system of two free fields: the radial
Higgs field $\varrho$ and the gauge field $\varepsilon$ whose masses are
$\sqrt 2$ and $\sqrt\xi$, respectively. Those are coupled at the boundary
site of the AHM ((\ref{hxy}) or (\ref{hpol})) each to its interacting
counterpart, {\it i.e.} $\varrho$ to $\rho$, and $\varepsilon$ to $\epsilon$.
Suppose for definiteness that the left boundary of the interacting system
separating it from the linear heat bath is at $j=0$ site of the chain. The
field equations at the boundary are then modified compared to
(\ref{eqmopol}), namely, the second and the fourth equation of (\ref{eqmopol})
are replaced by
\begin{eqnarray}
\dot b_{-1,0}&=&{1\over a}\left(\epsilon_{0,1}-2*\epsilon_{-1,0}\right)
-a\xi\epsilon_{-1,0}+D(\sqrt\xi,[\epsilon_{-1,0}])+F(\sqrt\xi,t),\nonumber\\
\dot\pi^\rho_0&=&{1\over a}\left(\rho_1\cos b_{0,1}-2\rho_0+1\right)
-2a\left(\rho_0-1\right)+D(\sqrt 2,[\rho_0-1])+F(\sqrt 2,t).\nonumber\\
\label{eqmo0}
\end{eqnarray}
In going from (\ref{eqmopol}) to (\ref{eqmo0}) we linearized
in the vicinity of $\rho_0=1, b_{-1,0}=0$ and, following Ref.~\cite{us},
introduced two terms describing interaction with the linear heat bath at
the boundary. The $D(m,[\sigma])$ term represents the reaction to the motion
of a boundary field $\sigma$ from the heat bath. The mass of the corresponding
heat-bath field is $m$. Explicitly,
\begin{equation}
D(m,[\sigma])=\int_{-\infty}^t \sigma(t')\chi_m(t-t')dt',
\label{dterm}\end{equation}
where the Fourier image of the causal response function $\chi_m(t)$ is
\begin{equation}
\tilde\chi_m(\omega)=
2i{\rm sign}(\omega)\sqrt{(\omega^2-m^2)(1+a^2(m^2-\omega^2)/4)}
\label{tildechim}
\end{equation}
for frequencies $m<|\omega|<\sqrt{m^2+4/a^2}$ and vanishes outside this range.
The $D(m,[\sigma])$ term thus describes dissipation processes: the choice of
$\chi_m(t)$ ensures that the waves traveling across the boundary into the
heat bath are completely absorbed. In order for the system to reach thermal
equilibrium with the heat bath, the $F(m,t)$ term describing thermal
fluctuations of a heat-bath field at the boundary should be included.
According to fluctuation-disspation theorem, $F(m,t)$ is a Gaussian random
variable whose time autocorrelation is related to $\chi_m(t)$:
\begin{equation}
\langle F(m,t)F(m,t+\tau)\rangle=\theta\int_0^\tau\chi_m(t')dt',
\label{fcorr}\end{equation}
where $\theta$ is the temperature.
We solve numerically the system of equations (\ref{eqmoxy}) together with
(\ref{eqmo0}) and its right-boundary analog. Numerical implementation of the
boundary heat bath is explained in detail in Ref. \cite{us}. As has already
been
mentioned, the Cartesian form of equations of motion in the bulk is used due to
its superior numerical properties. On the other hand, we use polar coordinates
to evolve boundary fields. In order to be able to transform from polar to
Cartesian coordinates at the boundary, we need to keep track of the
angular variable $\alpha$ of the boundary Higgs field. This is done with the
help of Gauss' law which for the linearized system gives
\begin{equation}
\dot\alpha_j={1\over a}\left(E_{j,j+1}-E_{j-1,j}\right).
\label{glawpol}\end{equation}
We use second-order Runge-Kutta algorithm for numerical solution of
(\ref{eqmoxy},\ref{eqmo0},\ref{glawpol}). A number of criteria were applied
in evaluating
the algorithm performance and in determining the value of the time step. In
particular, we tested the absorption properties of the simulated
heat bath by dropping the noise term from the equations of motion and cooling
an initially hot system. The resulting cooling curve was then compared to that
of the same system in a large real zero-temperature heat bath. In a similar
way we studied thermalization of an initially cold system in the heat bath.
The temperature was measured by averaging the kinetic energy of the radial
Higgs field over the system and over the time history. In all our simulations
the temperature of a thermalized system was found to be within 3\% of the
assigned value.

One special numerical issue to be dealt with in a real-time simulation of a
gauge theory is the accuracy of the Gauss' constraint. The equations we solve
are consistent with the Gauss' law. However, numerical errors give rise to a
small spurious static charge density which should be kept in check in order
to have no impact on the quantities we measure, in particular, the sphaleron
transition rate. Since the rate is exponentially sensitive to the sphaleron
energy, we computed the perturbative correction to the latter in presence of
a small static charge distribution $q(x)$, $\Delta E_{\rm sph}([q])$. The
corresponding expression is derived in the Appendix. We then used
$\beta \Delta E_{\rm sph}([q])$ averaged over the sphaleron positions as
a criterion of Gauss' law violation (here and in the following $\beta$ denotes
inverse temperature). In all our simulations the amount of the spurious
charge was far too small to have any measurable impact on the sphaleron
transition rate.

Our attention in this work is focused on the temperature dependence of the
rate. For this reason we performed all our simulations at a fixed value
of the gauge coupling $\xi=10$, with the exception of preliminary study at
$\xi=0.5$. Most of our measurements were done for a chain
of length $L=100$ and lattice spacing $a=0.5$. Since our results are to be
compared to the analytical prediction for AHM in the continuum, we checked
their dependence on both lattice cutoffs by performing additional simulations
at $a=0.25$ and at $L=50$. The Runge-Kutta time step was 0.01 for $a=0.5$
and 0.004 for $a=0.25$. For each set of $a$, $L$, and the temperature the
simulation time was $10^5$ time units.
To add confidence to our measurements we also performed microcanonical
simulations at a number of parameter values. The agreement with the canonical
results was good except for the $\beta=5, L=100, a=0.5$ case for which
the canonical simulation gives somewhat higher value of the rate.
\begin{figure}[t]
\epsfxsize 10.cm
\epsfbox{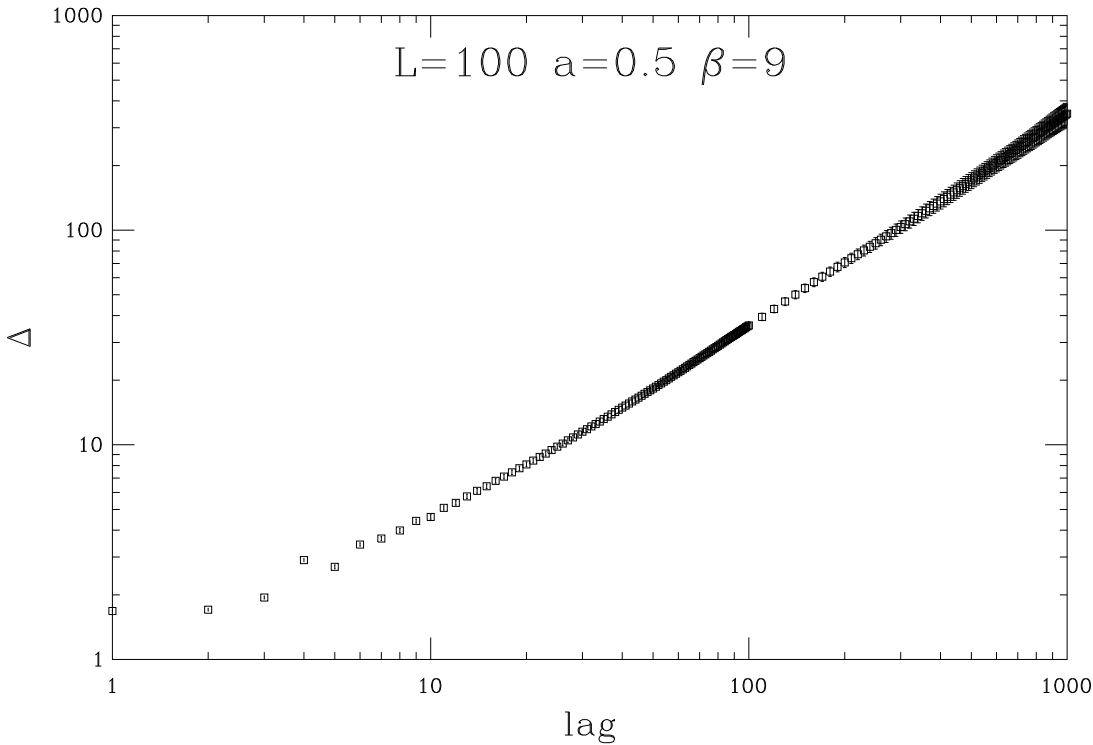}
\caption{Lag dependence of the average squared deviation of the Chern-Simons
variable.}
\label{dlag}
\end{figure}

Following Ref.~\cite{BdFH}, we extracted the sphaleron transition rate from
$\Delta_{\rm CS}(t)$, the time-averaged squared deviation of the
Chern-Simons variable $N_{\rm CS}\equiv(2\pi)^{-1}\int A(x)dx$ for a lag $t$.
For lags shorter than the average time between consecutive sphaleron
transitions
$\Delta_{\rm CS}(t)$ is determined by fluctuations of $N_{\rm CS}$ in the
vicinity of one of its vacuum values. At lags much longer than the lifetime
of a vacuum many uncorrelated sphaleron transitions would have occurred, each
changing the value of $N_{\rm CS}$ by an integer, and a random-walk behavior
sets in:
\begin{equation}
\Delta_{\rm CS}(t)=\Gamma L t,
\label{diffuse}\end{equation}
where $\Gamma$ is the sphaleron transition rate per unit length. Figure
\ref{dlag} illustrates the described lag dependence of $\Delta_{\rm CS}(t)$.
\begin{figure}[t]
\epsfxsize 10.cm
\epsfbox{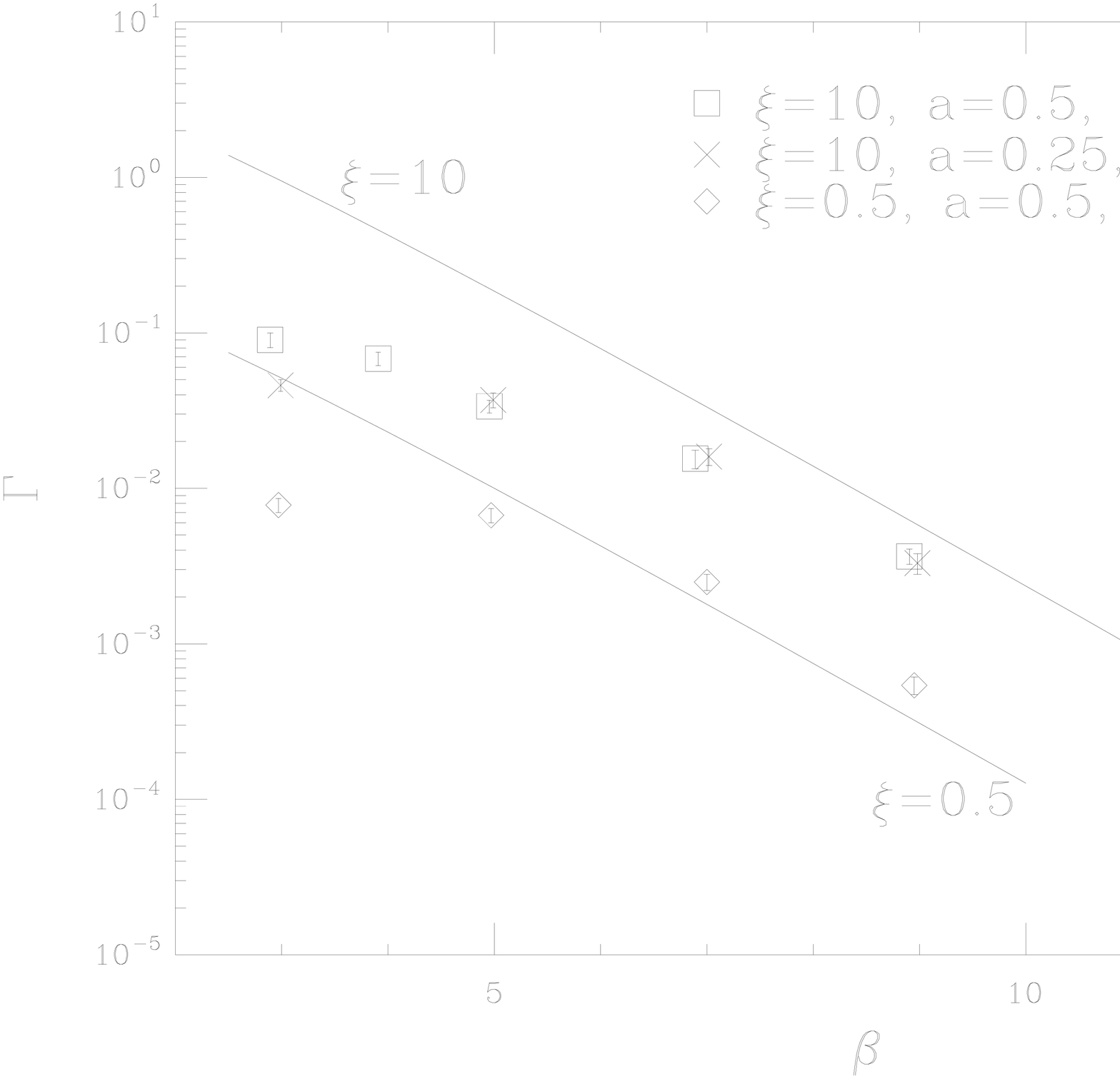}
\caption{Temperature dependence of the transition rate $\Gamma$.
The solid curves correspond to the TST prediction (13), with the value of
$\xi$ indicated near each curve.}
\label{rate}
\end{figure}

Once $\Delta_{\rm CS}(t)$ is known, it can be (at large enough values of $t$)
fitted to a straight line through the origin to yield $\Gamma$. Since the
values of $\Delta_{\rm CS}(t)$ at different $t$ are strongly correlated, we
found that, while the quality of fit remains high as more and more
$\Delta_{\rm CS}(t)$ data points are included, the error on $\Gamma$ is not
reduced significantly by fitting with more degrees of freedom. We therefore
simplified the procedure and extracted $\Gamma$ from a single value of
$\Delta_{\rm CS}(t)$ at $t=1000$. This choice of a lag is suitable since,
on one hand, it is much shorter than our total simulation time ($10^5$), while
on the other hand it is at least several times longer than the average time
between consecutive sphaleron transitions in the temperature range considered.
We also verified that $\Gamma$ remains constant in a wide range of lags
including the chosen one.
\begin{table}[t]
\centerline{\begin{tabular}{|l|l|l|l|l||l|l|l|l|l|} \hline
{$\xi$} & {$a$} & {$L$} & {$\beta$} & \multicolumn{1}{c||}{$\Gamma$} &
{$\xi$} & {$a$} & {$L$} & {$\beta$} & \multicolumn{1}{c|}{$\Gamma$} \\ \hline
\multicolumn{10}{|c|}{\rm RHB} \\ \hline
{10} & {0.5} & {100} & {2.89} & {$0.09\pm 0.01$} &
{10} & {0.5} & {50} & {8.87} & {$(37\pm 4)\times 10^{-4}$} \\
{10} & {0.5} & {100} & {3.91} & {$0.068\pm 0.007$} &
{10} & {0.25} & {50} & {2.99} & {$0.046\pm 0.004$} \\
{10} & {0.5} & {100} & {4.95} & {$0.034\pm 0.003$} &
{10} & {0.25} & {50} & {4.99} & {$0.037\pm 0.004$} \\
{10} & {0.5} & {100} & {6.89} & {$0.015\pm 0.002$} &
{10} & {0.25} & {50} & {7.02} & {$0.016\pm 0.002$} \\
{10} & {0.5} & {100} & {8.90} & {$(37\pm 4)\times 10^{-4}$} &
{10} & {0.25} & {50} & {8.98} & {$(33\pm 5)\times 10^{-4}$} \\
{10} & {0.5} & {100} & {11.85} & {$(32\pm 3)\times 10^{-5}$} &
{0.5} & {0.5} & {100} & {2.97} & {$(78\pm 8)\times 10^{-4}$} \\
{10} & {0.5} & {100} & {13.86} & {$(60\pm 7)\times 10^{-6}$} &
{0.5} & {0.5} & {100} & {4.97} & {$(67\pm 7)\times 10^{-4}$} \\
{10} & {0.5} & {50} & {4.96} & {$0.033\pm 0.005$} &
{0.5} & {0.5} & {100} & {7.00} & {$(25\pm 3)\times 10^{-4}$} \\
{10} & {0.5} & {50} & {6.88} & {$0.016\pm 0.002$} &
{0.5} & {0.5} & {100} & {8.95} & {$(54\pm 7)\times 10^{-5}$} \\ \hline
\multicolumn{10}{|c|}{\rm Microcanonical} \\ \hline
{10} & {0.5} & {100} & {4.94} & {$0.055\pm 0.007$} &
{10} & {0.25} & {50} & {3.00} & {$0.040\pm 0.004$} \\
{10} & {0.5} & {100} & {8.95} & {$(40\pm 5)\times 10^{-4}$} &
{} & {} & {} & {} & {} \\ \hline
\end{tabular}}
\caption{Summary of transition rate measurements. The inverse temperature
$\beta$ is given as deduced from the average kinetic energy of the radial Higgs
field.}
\label{tabrate}
\end{table}

Summary of all our rate measurements is presented in Table \ref{tabrate}, while
for Figure \ref{rate} we selected the results that best reflect the important
features of $\Gamma$ dependence on the inverse temperature $\beta$, as well as
on $a$.  Obviously, no measurable
dependence on $L$ is observed. The absence of finite-size effects
is to be expected of our heat-bath construction. Namely, at low temperatures
the linearized heat bath closely imitates the infinite extension of the
nonlinear system beyond the boundaries. At high temperatures the system becomes
less and less correlated in space and time, and, as a result, the boundary
effects lose importance. We also observe no dependence of the rate on $a$,
except for the highest temperature $(\beta=3)$ considered.
The virtual independence of $\Gamma$ of the lattice spacing at low temperatures
has been found in earlier work \cite{GRS,BdFH} and is confirmed by our
results. As will be shown shortly, the difference in the rates between the
$a=0.5$ and $a=0.25$ cases at $\beta=3$ is consistent with other properties
of the model at this temperature.
Our rate measurements are to be compared to the TST prediction \cite{BTs}
\begin{equation}
\Gamma={\omega_-\over{2\pi}}
\left({{6\beta E_{\rm sph}}\over{2\pi}}\right)^{1\over 2}
\left({{\Gamma(\alpha+s+1)\Gamma(\alpha-s)}
\over{\Gamma(\alpha+1)\Gamma(\alpha)}}\right)^{1\over 2}
\exp\left(-\beta E_{\rm sph}\right),
\label{TST}\end{equation}
where the sphaleron energy $E_{\rm sph}=2\sqrt{2}/3$, $\alpha^2=s(s+1)=2\xi$,
and $\omega_-^2=s+1$ is the negative squared eigenfrequency corresponding to
the sphaleron instability.

As Figure \ref{rate} clearly shows, the values of $\Gamma$ at low temperature
are close to those given by (\ref{TST}). While the approach of measured
$\Gamma$
to the TST prediction is slow, the two nearly coincide at the lowest
temperature
considered, $\beta=14$. This nice agreement shows once again that our heat bath
construction works as intended. But
the most notable feature of our results is their dramatic departure
from the TST-predicted values starting at about $\beta=5$. For $\xi=10$ the
discrepancy is a factor of 5 already at $\beta=5$, and grows at $\beta=3$
to a factor of 10 for $a=0.5$ and a factor of 20 for $a=0.25$. The situation
is similar for $\xi=0.5$. Moreover, in the latter two cases the rate
practically
does not grow between $\beta=5$ and $\beta=3$.  Note that the deviations are
much larger than our measurement error bars and are therefore statistically
significant.
\begin{figure}[t]
\epsfxsize 10.cm
\epsfbox{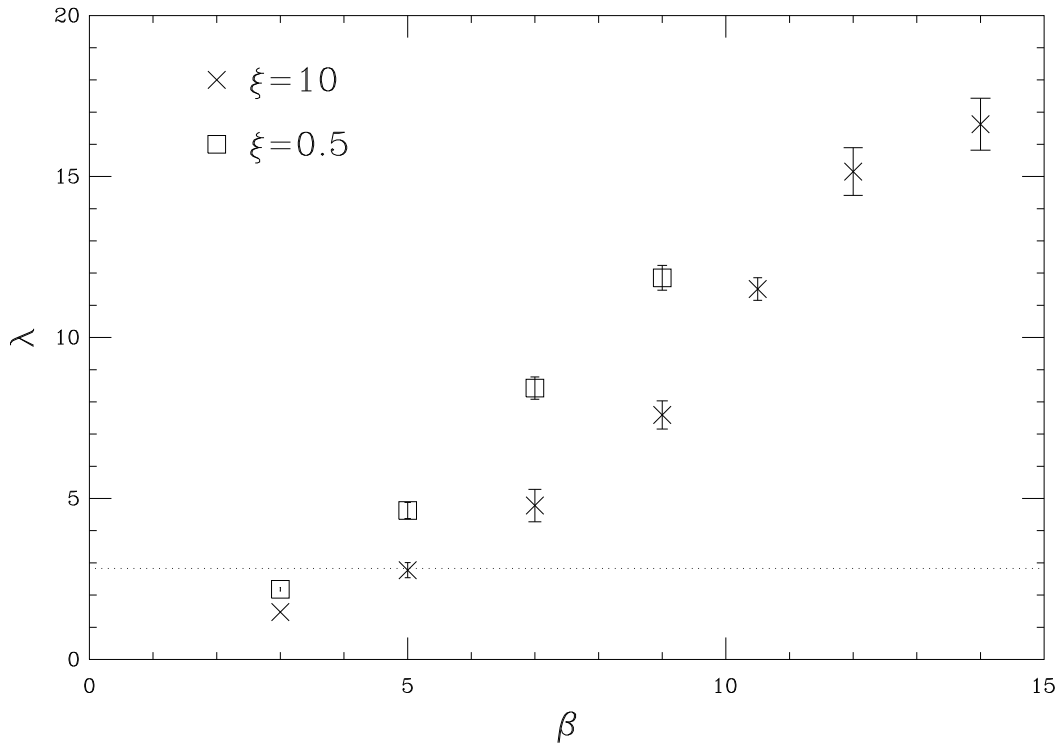}
\caption{Temperature dependence of the correlation length
deduced from the gauge-invariant two-point function (14). The sphaleron size
is shown by the dashed line for comparison.}
\label{lambda}
\end{figure}

It is natural to ask why this strong lagging of the measured rate behind the
TST one begins in the vicinity of $\beta=5$. To this end recall the underlying
assumption of (\ref{TST}): Chern-Simons number diffusion is dominated by
evolution of configurations resembling the vacuum into those resembling the
zero-temperature sphaleron. It is clear that the $N_{\rm CS}$ diffusion can
only be described in these terms as long as the Higgs field is correlated on a
length scale larger than the sphaleron size $(2\sqrt 2)$. The corresponding
correlation length $\lambda$ can be found by measuring a gauge-invariant
two-point function \cite{BF}
\begin{equation}
C_{jl}=\phi^*_j\phi_l\exp\left(-ia\sum_{k=j}^{l-1}A_{k,k+1}\right)
=\rho_j\rho_l\exp\left(-i\sum_{k=j}^{l-1}b_{k,k+1}\right).
\label{Cjl}\end{equation}
A rough estimate of $\lambda$ at low temperatures is obtained by averaging
$C_{jl}$ over the thermal ensemble with $H'$ of (\ref{hpol}) replaced by its
linearized version. Performing Gaussian integration over the $b$ variables one
finds $\lambda=2\beta$. This is, in fact, an overestimate of $\lambda$, since
thermal fluctuations of the radial Higgs field are not taken into account.
Figure \ref{lambda} shows the values of $\lambda$ obtained by fitting
$\langle C_{jl}\rangle$ to ${\rm const}\times\exp(-|j-l|/\lambda)$. As
expected,
the breakdown of the saddle-point approximation for the rate occurs as
$\lambda$
becomes smaller than the sphaleron size. Note that this is true for both values
of $\xi$ considered. At $\beta=3$ $\lambda=1.47$ for $\xi=10$, only about 3
times larger than the lattice spacing $a=0.5$. Hence the field strongly
fluctuates at length scales comparable to the lattice spacing. It is therefore
not surprising that we find the $a$ dependence of the rate at this temperature.

At this point it is unclear what causes the sharp slowdown of the rate growth
at the high-temperature end of our measurement range. We cannot exclude a
possibility that at temperatures in question crossing the
$N_{\rm CS}={\rm half-integer}$ separatrix in the configuration space of the
model \cite{Shaplat91} in close vicinity of the sphaleron saddle point is still
strongly preferred energetically, but is already suppressed entropically.
It is usually assumed \cite{Shaplat91,Amb3d,MW,Dine} that at temperatures above
the sphaleron energy the rate grows like a power of the temperature. It could
be that what we observe at $\beta\leq 5$ is a crossover from the exponential
to power-law behavior of the rate. It is not clear, however, that at
such a crossover the rate should stop growing as it does
for $\xi=10, a=0.25$ and for $\xi=0.5, a=0.5$. The only way to resolve this
puzzling situation is by rate measurements at still higher temperatures, as
well
as smaller lattice spacings. We plan to do so in the future.

To summarize, we performed an accurate measurement of the sphaleron transition
rate in AHM averaged over the canonical ensemble. The latter was obtained
by immersing the system in a realistic heat bath. The ergodicity of real-time
evolution was thus achieved without having to introduce an artificial viscosity
parameter. Our rate measurements approach the corresponding TST estimate
at low temperature. This is in agreement with Ref. \cite{GRS} where similar
measurements were performed microcanonically. The highest temperature studied
in
that work was $0.103 E_{\rm sph}$, well within the range of applicability of
TST.
In going beyond that range, we found dramatic slowdown of the rate growth, with
suppression factor as large as 20 relative to TST at the highest temperature
considered. Our measurements show that the breakdown of TST occurs as soon
as the correlation length deduced from $C_{jl}$ (\ref{Cjl})) becomes comparable
to the sphaleron size. This suggests that a similar object might serve as a
criterion for applicability of the sphaleron approximation in other theories,
including the realistic 3+1-dimensional case.

We gratefully acknowledge enlightening discussions with A.~I.~Bochkarev,
Ph.~de~Forcrand, E.~G.~Klepfish, A.~Kovner, and especially A.~Wipf.
Numerical simulations for this work were performed on the Cray YMP/464
supercomputer at ETH.

\section*{Appendix}
In this Appendix we give a perturbative estimate of the shift in the sphaleron
energy in presence of a small static charge density $q$. For simplicity we use
continuum,
rather than lattice, formulation of AHM. We shall also assume that the system
has an infinite length. Analogous to (\ref{hxy}), the
Hamiltonian depends on three pairs of canonical variables
$A, E, \rho, \pi_rho, \alpha, \pi_\alpha$. Static configurations are obtained
by minimizing the energy with respect to all the variables on a subspace
constrained by Gauss' law $\pi_\alpha=E'-q$ (prime means derivative with
respect
to the spatial variable $x$). Vacuum configurations correspond to the absolute
minimum of the energy on that subspace, while sphalerons minimize the energy
among configurations with $\rho=0$ at one point. Two of the variables, $A$ and
$\alpha$, enter the Hamiltonian only in combination $b=\alpha'-A$, thereby
effectively reducing the number of variables by one. It is convenient to
define a new complex field
\begin{equation}
\Phi(x)=\rho(x)\exp\left(i\int_{-\infty}^x b(x')dx'\right).
\label{Phi}\end{equation}
After eliminating $\pi_\alpha$ with the help of Gauss' law one obtains
\begin{equation}
H={1\over 2}\int dx \left[\xi E^2+\pi_\rho^2+{{(E'-q)^2}\over{|\Phi|^2}}+
|\Phi'|^2 +{1\over 2}(|\Phi|^2-1)^2\right],\label{hcont}
\end{equation}
Extremization of
energy leads, apart from the trivial condition $\pi_\rho=0$, to a system
of coupled equations for $\Phi$ and $E$:
\begin{eqnarray}
\xi E-\left[{{E'-q}\over{|\Phi|^2}}\right]'=0,
\Phi''+{{(E'-q)^2\Phi}\over{|\Phi|^4}}-\Phi(|\Phi|^2-1)=0.\label{mineq}
\end{eqnarray}
These equations are simplified if we introduce an electrostatic potential for
the electric field: $E=Y'$. If we require that $q$ vanishes for $|x|$ above
certain value, $E$ and $|\Phi|$ must respectively approach 0 and 1 as
$x\rightarrow\pm\infty$. The electrostatic potential $Y$ will then approach
a constant value. If that constant is chosen to be 0, (\ref{mineq}) takes form
\begin{eqnarray}
Y''-\xi|\Phi|^2Y=q;
\Phi''+\xi^2Y^2\Phi-\Phi(|\Phi|^2-1)=0.\label{mineqY}
\end{eqnarray}
We are interested in perturbative corrections to the lowest
order in $q$ to the vacuum $Y=0, \Phi=1$ and to the sphaleron configuration
$Y=0, \Phi=\tanh(x/\sqrt 2)$. We will show in the following that in both
cases the correction to $Y$ is first order in $q$. It is then clear from
the second equation of (\ref{mineqY}) that the correction to $\Phi$ is at
best second order. Inspecting the Hamiltonian (\ref{hcont}) and bearing in
mind that the unperturbed solutions extremize $H$ for $q=0$ we conclude that
the correction to the energy is second order in $q$ and comes solely from
the first-order correction to $Y$. Our task therefore reduces to solving the
first equation of (\ref{mineqY}) in the unperturbed $\Phi$ background. Writing
$Y(x)$ as $\int G(x,y)q(y)dy$ we obtain an equation for the Green's function
$G(x,y)$:
\begin{equation}
\left(\partial^2_x-\xi|\Phi(x)|^2\right)G(x,y)=\delta(x-y).\label{eqG}
\end{equation}
It is easy to verify that
\begin{equation}
G_v(x,y)=-{1\over{2\sqrt\xi}}\exp\left(-{\sqrt\xi}|x-y|\right)\label{Gv}
\end{equation}
for the vacuum ($\Phi(x)=1$). For the sphaleron  configuration
($\Phi(x)=\tanh(x/\sqrt 2)$) the solution of (\ref{eqG}) is also
straightforward
but somewhat cumbersome. The result can be expressed in terms of
hypergeometric functions:
\begin{eqnarray}
G_s(x,y)&=&-C\cosh^{-\sqrt{2\xi}}\left(x/\sqrt 2\right)
\cosh^{-\sqrt{2\xi}}\left(y/\sqrt 2\right)\nonumber\\
&&\times F\left(\alpha_+,\alpha_-;\gamma;
{\exp\left(x/\sqrt 2\right)\over{2\cosh\left(x/\sqrt 2\right)}}\right)
\nonumber\\
&&\times F\left(\alpha_+,\alpha_-;\gamma;
{\exp\left(-y/\sqrt 2\right)\over{2\cosh\left(y/\sqrt 2\right)}}\right)
\label{monster}\end{eqnarray}
if $y\geq x$, with $x$ and $y$ interchanged otherwise. Here
\begin{eqnarray}
\alpha_\pm&=&{{1+\sqrt{8\xi}\pm\sqrt{1+8\xi}}\over 2};\
\gamma={{1+\alpha_++\alpha_-}\over 2};\nonumber\\
C&=&{{\Gamma(\alpha_+)\Gamma(\alpha_-)}
\over{4^{1+\sqrt{2\xi}}\Gamma(\gamma)\Gamma(\sqrt{2\xi})}}.\end{eqnarray}
Substituting the correction to electric field into (\ref{hcont}) we find the
energy shift of a state due to the static charge $q$:
\begin{equation}
\Delta H=-{\xi\over 2}\int dxdyG(x,y)q(x)q(y).
\end{equation}
This result is hardly surprising: $G(x,y)$ is nothing but the Coulomb potential
at $x$ due to a unit charge at $y$. Therefore to the lowest order in $q$ the
energy shift is equal to the Coulomb energy of external charge distribution.
The correction to the sphaleron energy barrier then follows immediately:
\begin{equation}
\Delta E_{\rm sph}=-{\xi\over 2}\int
dxdy\left(G_s(x,y)-G_v(x,y)\right)q(x)q(y).
\label{desph}\end{equation}

\end{document}